\documentstyle[epsfig]{mn}

\title[DNOs and QPOs in Cataclysmic Variables: VIII.]
{Dwarf Nova Oscillations and Quasi-Periodic Oscillations in Cataclysmic Variables, VIII. 
VW Hyi in outburst observed with the Southern African Large Telescope\thanks{based
on observations made with the Southern African Large Telescope (SALT)}}
\author[Patrick A. Woudt et al.]
       {P.A.~Woudt$^1$\thanks{email: Patrick.Woudt@uct.ac.za},
        B.~Warner$^{1,2}$, D.~O'Donoghue$^3$, D.A.H.~Buckley$^3$, M.~Still$^{3,4}$ \newauthor
        E.~Romero-Colemero$^3$, P.~V\"ais\"anen$^3$\\
        $^1$ Department of Astronomy, University of Cape Town, Private Bag X3,
        Rondebosch 7701, South Africa\\
        $^2$ School of Physics and Astronomy, Southampton University, Highfield, 
        Southampton SO17 1BJ, UK\\
        $^3$ South African Astronomical Observatory, PO Box 9, Observatory 7935, South Africa\\
        $^4$ NASA Ames Research Center, Moffett Field, CA 94035, USA}
\date{2009 April 28}

\begin{document}

\maketitle

\begin{abstract}
     We analyse four light curves obtained at high time resolution ($\sim$ 0.1 s) with 
the 11-m Southern African Large Telescope, at the ends of two normal outbursts and one 
superoutburst of the dwarf nova VW Hyi. All of these contain at least some 
Dwarf Nova Oscillations (DNOs), which, when at their highest amplitudes, are 
seen in unprecedented detail. In addition to the expected DNOs with periods $>$ 20 s 
we find a previously unknown modulation at 13.39 s, but none at shorter periods. 
The various DNOs and their interaction with the longer period Quasi-periodic 
Oscillations are interpreted in terms of the model of magnetically controlled 
flow from an accretion disc proposed earlier in this series of papers. Our 
observations include rare DNOs very late in outburst; we find that the 
fundamental period does not increase beyond $\sim$ 90 s, which is the same 
value that the independent ``longer period DNOs'' converge on.

\end{abstract}

\begin{keywords}
accretion, accretion discs -- binaries: close, dwarf novae, cataclysmic variables -- stars: oscillations --
stars: individual: VW Hyi 
\end{keywords}

\section{Introduction}

     This series of papers is investigating the optical modulation phenomena observed 
in cataclysmic variable stars (CVs) at moderate time resolution, in particular the 
Dwarf Nova and Quasi-periodic Oscillations (DNOs and QPOs) that are associated with 
accretion from a companion star through an accretion disc onto the surface of the 
white dwarf primary. A review of the general properties of CVs is given in 
Warner (1995). In previous papers (Woudt \& Warner 2002: hereafter Paper I; 
Warner \& Woudt 2002: hereafter Paper II; Warner, Woudt \& Pretorius 2003: hereafter 
Paper III; Warner \& Woudt 2006: hereafter Paper IV; Pretorius, Warner \& Woudt 2006: 
hereafter Paper V, and Warner \& Pretorius 2008: hereafter Paper VI) we have used 
integration times of typically $\sim$ 5 s to study in detail  the DNOs (periods in 
the range 3 -- 40 s) and QPOs (two kinds; ones with periods $\sim$ 16 times the DNO 
periods and others of much longer periods, typically 1000 -- 2000 s), and have 
discovered the existence of an intermediate class of oscillations (long period DNOs: 
lpDNOs, periods $\sim$ 4 times those of the DNOs). A review of the rich phenomenology 
of these oscillations, which includes systematic variations of period with rate of 
mass transfer ($\dot{M}$) through the disc and the interactions of the various types 
of oscillation, is given in Warner (2004), which also presents an interpretation of 
some aspects in terms of a magnetic accretion model, proposed initially by 
Paczynski (1978) and developed further in Papers II, V and VI. In many respects
it is a generalisation of the standard model of intermediate polars. One aspect that 
we need to mention immediately is that most DNOs are single valued and interpreted 
as being generated by magnetically-controlled accretion onto a freely moving equatorial accretion 
belt on the white dwarf primary. This DNO we refer to as having a sidereal period, $P_{\rm DNO}$, 
or frequency $\omega_{\rm DNO}$. Occasionally a second DNO appears (often with a strong 
first harmonic), at what we call the synodic periodic $P_{\rm SYN}$ because the difference 
in frequency between the two DNOs is equal to the frequency $\omega_{\rm QPO}$ of the 
QPO (when present), i.e., $\omega_{\rm DNO} - \omega_{\rm SYN} = \omega_{\rm QPO}$. The 
basic model is that the beam of high energy radiation arising from the accretion 
zone on the equatorial belt is reprocessed off a traveling wave near the inner edge of
the magnetically truncated disc that is the source of the observed QPOs (Paper II).

\begin{table*}
 \centering
  \caption{The observing log of observations of VW Hyi taken with the Southern African Large Telescope.}
  \begin{tabular}{@{}lclccllccl@{}}
   Run        & Filter  & Date of first obs.& UTC start  & HJD start & Outburst     & Length & $t_{in}$ & $T$  & V \\
              &         & (start of night)  & (hh:mm:ss.ss)    & (+ 245\,0000.0)   & type         & (h) & (s)    & (d) & (mag)\\[10pt]
 SALT004 & U  & 2005 December 09 &  21:39:32.74   & 3714.40199535 & Normal        &  0.2, 0.95  &  0.080 & 0.10 $\rightarrow$ 0.20 & 12.4 \\
 SALT005 & -- & 2006 December 18 &  20:09:02.64   & 4088.33904461 & Normal        &  1.89       &  0.080 & 1.74 $\rightarrow$ 1.83 & 13.4 \\
 SALT016 & -- & 2008 January 05  &  19:54:55.72   & 4471.32895922 & Super         &  2.38       &  0.120 & --0.15: $\rightarrow$ --0.05: & 11.8 \\
 SALT017 & -- & 2008 January 06  &  19:10:33.67   & 4472.29813736 & Super         &  0.55       &  0.120 & 0.82: $\rightarrow$ 0.84: & 13.0 \\
\end{tabular}
{\footnotesize 
\newline 
Notes: The time $T = 0$ is based on a template light curve of VW Hyi as defined in Paper I (see figure 5 of Paper I); $t_{in}$ is the integration time.\\
}
\label{dno7tab1}
\end{table*}

   Most CVs are quite faint and therefore the study of their rapid brightness 
oscillations is often severely photon-limited, with individual pulse shapes not fully
resolved. Although a few moderate time 
resolution photometric and spectroscopic studies of CVs have been made with 
8-m class telescopes (e.g., V2051 Oph: Steeghs et al. 2001), 
there has hitherto not been a substantial study of the 
short period modulations made at considerably higher time resolution and good signal-to-noise. 
In particular, for the unpredictable outbursts of dwarf novae there will in general be 
difficulties of obtaining observing opportunities at the crucial times. 
Here, in what marks a shift in this series of papers to ultra-high  
time-domain studies of DNOs and QPOs with 10-m class facilities, 
we report the results from photometric runs with $\sim$ 0.1 s integrations made 
with the Southern African Large Telescope (SALT) in `slot-mode' during its start-up phase, where 
rapid access to the telescope was shown to be feasible. The telescope and its 
photometer are described in O'Donoghue et al.~(2006). Most observations
were taken in white light (unfiltered), except for the 2005 December 9 observing
run when a $U$-band filter was used.

     In Section 2 we give details of the observing runs and a general 
description of the slow and rapid brightness variations directly observable 
in the light curves; in Section 3 we give the results of detailed analysis 
of the light curves. Section 4 provides a brief synopsis.

\section{Observation}

\subsection{The observational procedure}

\begin{figure}
\centerline{\hbox{\psfig{figure=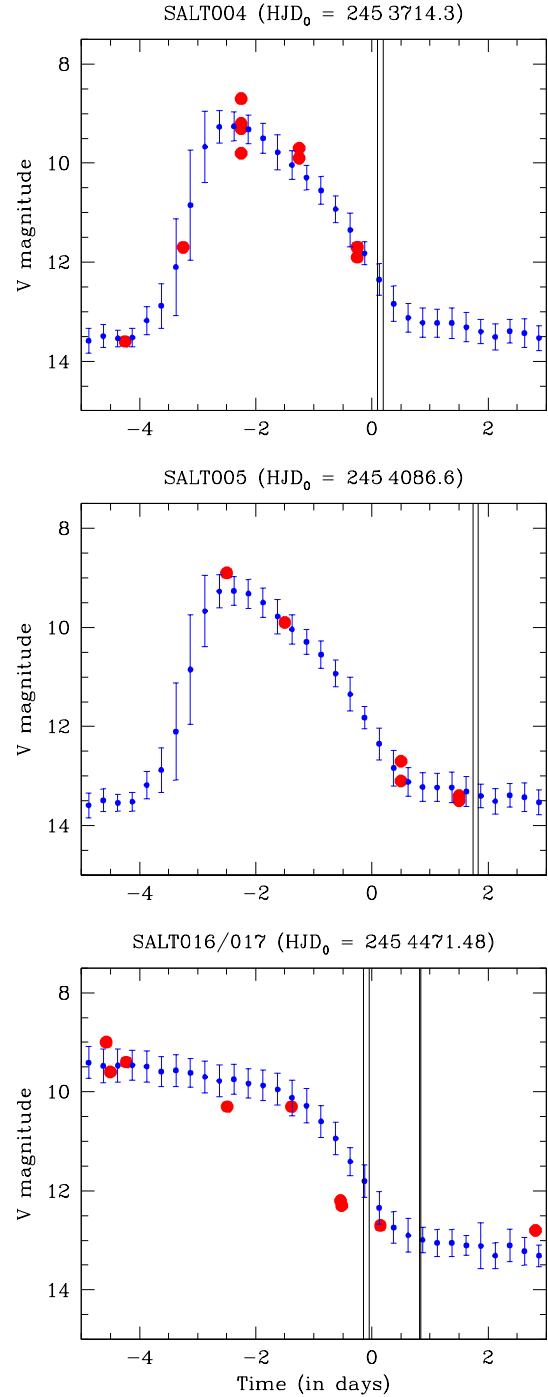,width=8.4cm}}}
  \caption{The outburst profiles of the December 2005 (SALT004), December 2006 (SALT005)
and January 2008 (SALT016/017) outbursts of VW Hyi. The large filled circles represent observations
gathered by observers from the Variable Star Section of the RASNZ (December 2005 and 2006) and the AAVSO
(January 2008); the small dots with error bars correspond to the outburst templates (Paper I). The vertical
lines show the time period of SALT observations relative to the outburst template.}
 \label{templates}
\end{figure}

Details of the VW Hyi observations are given in Table~\ref{dno7tab1}. The parameter $T$ is the time in 
days derived from comparison of the outburst light curve (obtained from the American 
Association of Variable Star Observers (AAVSO) and/or the Variable Star Section of the 
Royal Astronomical Society of New Zealand (RASNZ)) with the appropriate outburst `template' 
(figure 5 of Paper I), on which $T$ = 0 is defined (it is when the V magnitude passes 
through 12.5 on the decline). This enables the evolution of the outburst decline 
light curves to be aligned, see Fig.~\ref{templates}. For the 2008 January outburst the sampling of the outburst
light curve was sparse over the decline phase; from the relative magnitude difference of
VW Hyi on 2008 January 5 and January 6, derived from our SALT photometry (using the same 
reference star), an appropriate match with the outburst template was found.
As it was not possible to put the SALT observations 
onto a magnitude scale we have used the templates to estimate V magnitudes at the 
times of observation.

     The times of observation were chosen to catch the complex DNO behaviour that 
occurs near the end of outburst, during the final transition to quiescence, where frequency 
doubling and tripling are seen (Paper IV).

\subsection{Data reduction}

All standard photometric data reductions were performed in IRAF\footnote{IRAF is 
distributed by the National Optical Astronomy Observatory, which is operated by the 
Association of Universities for Research in Astronomy, Inc., under cooperative agreement 
with the National Science Foundation.}, including gain corrections of the different 
amplifiers (SALTICAM has two CCDs with two amplifiers each), merging of the different amplifiers, a 
superflat correction to remove the effects of vignetting by the spherical abberation
corrector in `slot-mode' observations, and aperture photometry using \textsc{DAOPHOT}.

Relative photometry was obtained for all runs by including a reference star in the field of
view. This effectively removes the effects of the moving pupil and varying aperture of SALT 
during the course of a photometric run.

\subsection{Preliminary description of the nature of the light curves}

\begin{figure}
\centerline{\hbox{\psfig{figure=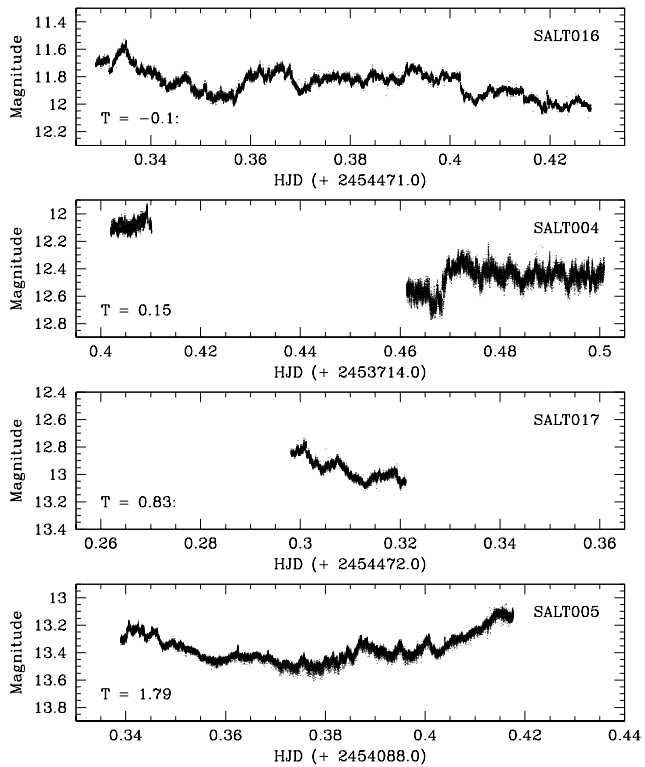,width=8.4cm}}}
  \caption{The four light curves obtained with the Southern African Large Telescope. From top to bottom,
they are presented in order of increasing $T$, where $T$ represents time in days along the decline from
outburst, defined from an outburst `template' light curve (Paper I).}
 \label{lightcurves}
\end{figure}

The light curves for the four runs are shown in order of $T$ and in compacted 
form in Fig.~\ref{lightcurves}. In the 9 December 2005 run the large gap is caused by an 
interruption of observation, during which VW Hyi decreased in brightness by $\sim$ 0.4 mag. 
There is a noticeable change in the amplitude of rapid variations from run to run. 
This is almost entirely due to the differing prominence of DNOs. 

\subsubsection{2008 January 5 (SALT016)}

This lightcurve, at the end of a superoutburst, is irregular and no DNOs are 
directly visible in it -- but the detailed analysis given below shows that DNOs at 
very low amplitude are actually occasionally present. Short trains of QPOs are sometimes 
visible. From the AAVSO archive there are only two magnitude estimates available for 
the final descending part of this superoutburst, which makes estimation of $T$ uncertain. 
However, there are independent indicators that lead to the values quoted in Table~\ref{dno7tab1}. 
The change of mean magnitude between this and the next night shows that VW Hyi was on a 
rapidly descending part of the outburst lightcurve. The DNOs that appear in the second 
half of the 5 January run have a period $\sim$ 22 s, which is at the beginning of 
the steep climb in period seen in figure 1 of Paper IV, and shows that we have for 
the first time caught the transition from rare to frequent DNOs at the start of 
the rapid deceleration. In fact, we have not seen DNOs at this stage of a superoutburst 
before (e.g. run S1594 in table 1 of Paper IV). These properties indicate $T$ $\sim$ 0.0.   

Towards the end of the run there are 24.8 s DNOs and 400 s QPOs, which have a ratio 
of 16.1, close to the expected relationship.

\begin{table*}
 \centering
  \caption{DNOs and QPOs in VW Hyi during decline from normal and super outburst. The runs have been
subdivided into sections of different length, for each of which the mean time $<T>$ is given.}
  \begin{tabular}{@{}lccccclc@{}}
 Run No.   & $<T>$ & Length         & \multicolumn{3}{c}{DNOs}  &  \multicolumn{2}{c}{QPOs} \\
           & (d)       & (s)            & \multicolumn{3}{c}{(periods in seconds)} & \multicolumn{2}{c}{(period in seconds)}\\
           &           &                & \multicolumn{3}{c}{[amplitude in mmag]} & \multicolumn{2}{c}{[amplitude in mmag]} \\[2pt]
           &           &                & Fundamental                   & First harmonic        & Second harmonic &  &  \\[8pt]
SALT004    & \hfill 0.11     & \hfill  706           & 25.64 $\pm$ 0.01 \hfill [26.6] & \hspace{1.1cm}    -- \hspace{1.1cm}   & \hspace{1.1cm}   -- \hspace{1.1cm}   & &  -- \\
           & \hfill  0.17     & \hfill 1659          & 27.65 $\pm$ 0.01 \hfill [23.3] &     --    &    --    &       &  --  \\
           & \hfill  0.18     & \hfill 648           & 28.49 $\pm$ 0.02 \hfill [25.5] &     --    &    --    &\vline &    \\
           & \hfill  0.19     & \hfill 657           & 28.76 $\pm$ 0.04 \hfill [15.0] &     --    &    --    &\vline &  403 $\pm$ 1 [28.4] \\
           & \hfill  0.20     & \hfill 812           & 28.82 $\pm$ 0.02 \hfill [19.5] &     --    &    --    &\vline &    \\[5pt]
SALT005    & \hfill  1.74     & \hfill 708           &            --                  &     --    &  28.82 $\pm$ 0.04 \hfill [7.6]  & &  -- \\
           & \hfill  1.76     & \hfill 734           & 91.43 $\pm$ 0.37 \hfill  [4.0] &  45.69 $\pm$ 0.08 \hfill [4.7]  &              --                  & &  -- \\
           & \hfill  1.77     & \hfill 432           &            --                  &  42.14 $\pm$ 0.17 \hfill [5.9]  &              --                  & &  -- \\
           & \hfill  1.79     & \hfill 605           &              --                &               --                &  28.93 $\pm$ 0.05 \hfill [6.4]   &\vline &     \\
           & \hfill  1.80     & \hfill 605           &              --                &  45.30 $\pm$ 0.09 \hfill [8.8]  &              --                  &\vline & 424 $\pm$ 1 [19.7]  \\
           & \hfill  1.81     & \hfill 432           &              --                &  41.25 $\pm$ 0.16 \hfill [4.5]  &              --                  &\vline &     \\
           & \hfill  1.81     & \hfill 544           &              --                &  45.09 $\pm$ 0.13 \hfill [6.5]  &  28.49 $\pm$ 0.06 \hfill [6.1]   &\vline &     \\[5pt]
SALT016    & \hfill --0.09    & \hfill 845           & 22.05 $\pm$ 0.03 \hfill [4.6]  &     --    &               --                & &  -- \\
           & \hfill --0.05    & \hfill 1146          & 24.78 $\pm$ 0.03 \hfill [3.1]  &     --    &               --                & & 400 $\pm$ 1 [25.7] \\[5pt]
SALT017    & \hfill  0.84     & \hfill 1987          &            --                  &  38.33 $\pm$ 0.03 \hfill [4.4]  &               --                & &  -- \\[5pt]
\end{tabular}
{\footnotesize 
\newline 
 }
\label{dno7tab2}
\end{table*}

\subsubsection{2005 December 9 (SALT004)}

In the rapid variation regime the immediately obvious features of the light curve (Fig.~\ref{lightcurves})
are the presence of persistent large amplitude ($\sim$ 0.1 mag) DNOs and several 
cycles of similar amplitude QPOs. A Fourier transform (FT) of the entire longer section 
of the run provides an average period of $\sim$ 28.4 s for the DNOs, and mean period 
of $\sim$ 403 s for the QPO, though the individual cycles 
have periods that range over $\sim$ 380 -- 440 s, 
demonstrating their quasi-periodicity. We will look at these modulations in greater 
detail below. We note here that the ratio $P_{\rm QPO}/P_{\rm DNO}$ = 403/28.4 = 14.2 is similar 
to what we have found before, and that the two periods fit perfectly onto the correlation 
diagrams of $P_{\rm DNO}$ and $P_{\rm QPO}$ with $T$ (figures 1 and 10 of Paper IV).  

\begin{figure}
\centerline{\hbox{\psfig{figure=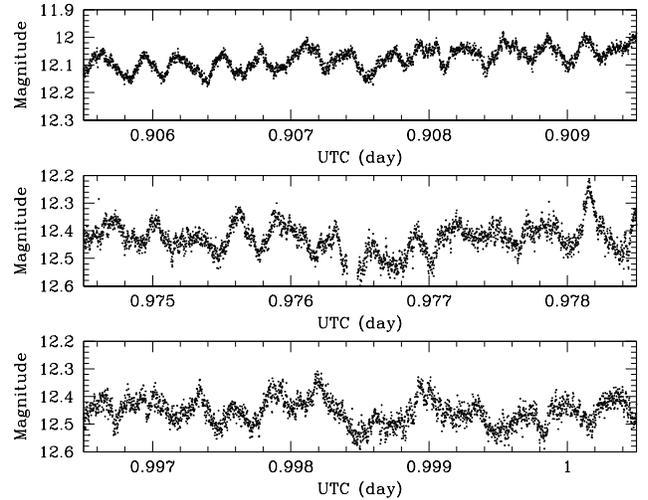,width=8.4cm}}}
  \caption{Three short sections of the light curve of VW Hyi (run SALT004) obtained with the Southern African
Large Telescope at 80 ms time resolution.}
 \label{SALT004lc1}
\end{figure}

     In Fig.~\ref{SALT004lc1} we show some examples (each 346 s long) 
of magnified short sections of the light curve, 
which demonstrate the good signal/noise aspects of the light curve -- in fact, for 
study of the DNOs there is considerable over-sampling, but the potential for 
finding higher frequency modulations is evident, which was largely the stimulus for 
carrying out this investigation. Although it emerges that there are no new periodicities 
to be found with periods $<$ 10 s, the short sections of DNO modulation shown in 
Fig.~\ref{SALT004lc1} shows that the maxima and minima of some cycles are very sharp, with significant 
changes on time scales $<$ 1 s. Previous observations would have integrated over 
these rapid changes.

\subsubsection{2008 January 6 (SALT017)}

This very short run shows three obvious cycles of a long time scale modulation, 
probably a QPO. There are no DNOs immediately evident in the light curve.

\subsubsection{2006 December 18 (SALT005)}

The overall curvature of the time series data is not understood -- maximum brightness of the 
orbital hump is predicted (van Amerongen et al.~1987) to be roughly in the centre 
of the run, when VW Hyi was observed at its lowest brightness. DNOs and QPOs are 
occasionally present in the light curve, which is rare with VW Hyi essentially back 
in quiescence, but not unprecedented -- the light curve obtained on 14 Nov 1985 (Paper I) 
at $T$ = 2.26 had a DNO in it with $P_{\rm DNO}$ = 24.7 s, which seemed anomalous at the time 
but following the discussion in Paper IV on frequency doubling and tripling we now recognize 
it as a second harmonic, not the fundamental DNO.

\section{The DNOs and QPOs}

We now look in more detail at the DNO and QPO behaviours in each of the four light curves. 
None of the FTs shows any coherent periodicities at periods shorter than 13 s.

\subsection{The 2008 January 5 Run (SALT016)}

We divided the light curve into 10 equal subsections (of length $\sim$ 600 s), 
the first six of which contain no evidence for DNOs, after which a clear signal 
at 22.05 s is seen in the seventh section. The next obvious DNO in the FT occurs in 
the combined ninth and tenth sections at a period of 24.78 s and also a very strong 
QPO at 400 s. More details are given in Table~\ref{dno7tab2}.

\begin{figure}
\centerline{\hbox{\psfig{figure=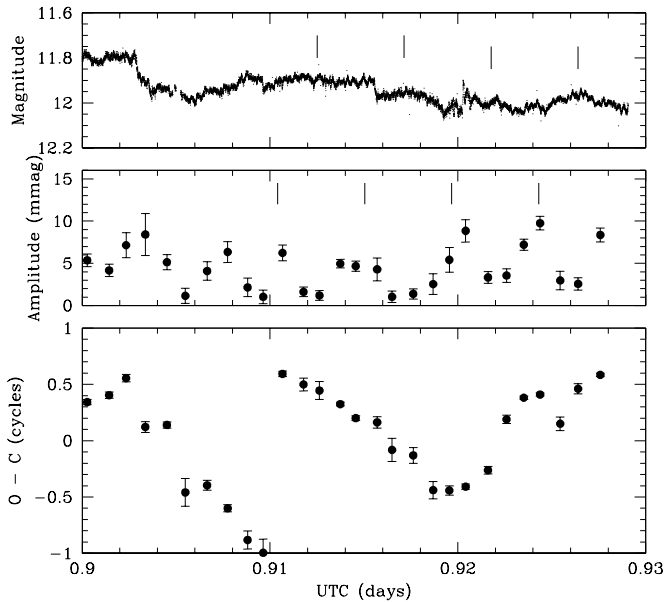,width=8.8cm}}}
  \caption{The $A - \phi$ diagrams of the last section of run SALT016 for the
24.78-s DNO. Each dot represents $\sim$ 5 DNO cycles with 50\% overlap between them. The top panel 
shows the corresponding light curve with the 400-s QPO marked by the vertical bars. The bars in
the central panel mark peaks in amplitude, spaced at the 400-s QPO.}
 \label{SALT016omc1}
\end{figure}

     We now introduce an $A - \phi$ diagram, as defined in Paper IV, which gives 
the Observed -- Calculated amplitude and phase variations relative to a fixed period. 
The $A - \phi$ diagram of the final part (UTC 0.90 -- 0.93) of the light curve, relative to the 
period 24.78 s, is shown in Fig.~\ref{SALT016omc1}. Up until 0.918 the DNO has a constant period, after 
which the period lengthens, as part of the general increase in period normally seen 
in the decline phase of the light curve. During this latter part there is strong 
amplitude modulation (middle panel), seen to be approximately in antiphase with the 
QPO in the light curve (upper panel). This is an indication that we are seeing the 
synodic and not the sidereal DNO -- the DNO rotating beamed radiation is reprocessed 
off the 400 s QPO travelling wave. This implies $P_{\rm DNO}$ = 23.33 s.

\subsection{The 2005 December 9 Run (SALT004)}

We divided the run into five subsections. The FTs naturally show very strong DNO peaks, 
their periods increasing through the run, with a mean period of 25.64 s and mean 
amplitude 26.6 mmag in the short first section and periods at 27.65 s (23.3 mmag) in 
the second section, increasing to 28.82 s (19.5 mmag) towards the end. 
There are short lived appearances of lpDNOs, e.g. at 86.6 s and 94.8 s. This general 
behaviour only serves to add more weight to the same correlations previously found (figure 3 
of Paper I and figure 1 of Paper IV).

\begin{figure}
\centerline{\hbox{\psfig{figure=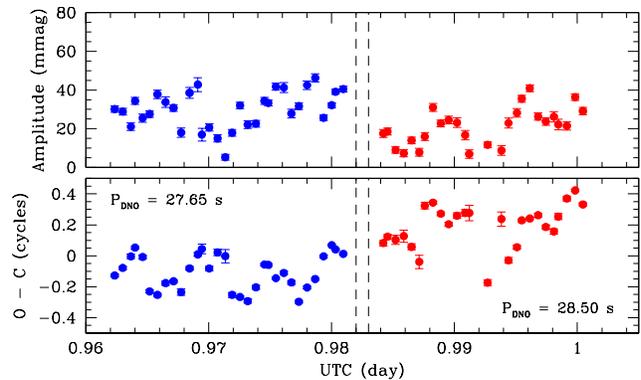,width=8.4cm}}}
  \caption{The A -- $\phi$ diagrams for the second part of the SALT004 light curve, relative
to a DNO at 27.65 s (left panels), and a DNO at 28.50 s (right panels), respectively. Each dot
represents $\sim$ 2.5 DNO cycles; there is a 50\% overlap.}
 \label{SALT004omc1}
\end{figure}

    Fig.~\ref{SALT004omc1} shows the $A - \phi$ diagram for the second part 
of the light curve (subsections 2 to 5), 
here split into only two sections, which shows a sudden change of behaviour roughly midway 
through, where the mean DNO period (adopted for calculating $\phi$) changes from 27.65 s 
to 28.50 s. The first section shows cyclical phase variations (Fig.~\ref{SALT004omc1}), 
an FT of which gives a period of 475 $\pm$ 13 s, but an FT of the same piece of light curve 
itself shows no significant QPO period demonstrating (as occasionally seen before) that 
although the rotating DNO beam intersects the QPO travelling wave, the latter is not
necessarily visible in the direction of the observer.

\begin{figure}
\centerline{\hbox{\psfig{figure=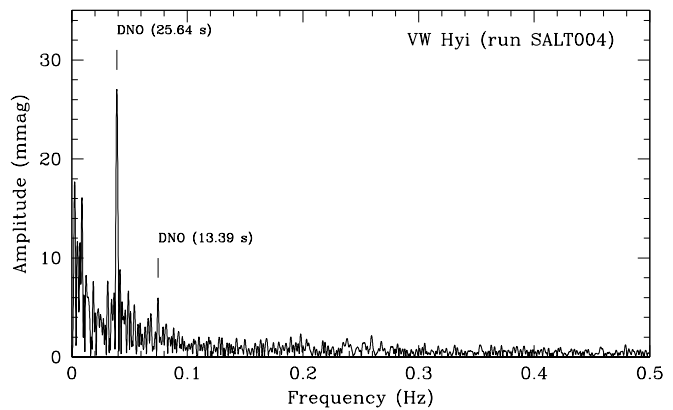,width=8.8cm}}}
  \caption{The Fourier transform of the first section of run SALT004.}
 \label{SALT004ft1}
\end{figure}

      The FT (Fig.~\ref{SALT004ft1}) of the short first part of the run shows a peak at 13.39 s, which
 has not been previously seen in any of the extensive studies of VW Hyi. We see from 
figure 10 of Paper IV that the expected QPO fundamental at $T$ = 0.10 -- 0.20 d 
is $\sim$ 400 -- 550 s with a first harmonic probably appearing during this at $\sim$ 300 s. 
The cyclic phase variation seen in the short first part of the run corresponds to the 
QPO fundamental. The 13.39 s modulation is not the harmonic of the observed DNO 
(the sidereal period at 25.64 s), but if it is the first harmonic of the synodic 
period (i.e., $P_{\rm SYN}$/2) then the beat between sideral and synodic periods is at 602 s, which 
is in the neighbourhood of the QPO period expected from the general correlation with $T$. 
In the first section of the second, long part of the run, with VW Hyi having faded by 
0.4 mag, the main DNO modulation is at 27.45 s and the short period modulation has increased 
in period to 15.15 s. If it is still a synodic first harmonic it would produce an 
expected beat at 292 s, which presumably is now the first harmonic of 
the (still unobservable) QPO.

\begin{figure}
\centerline{\hbox{\psfig{figure=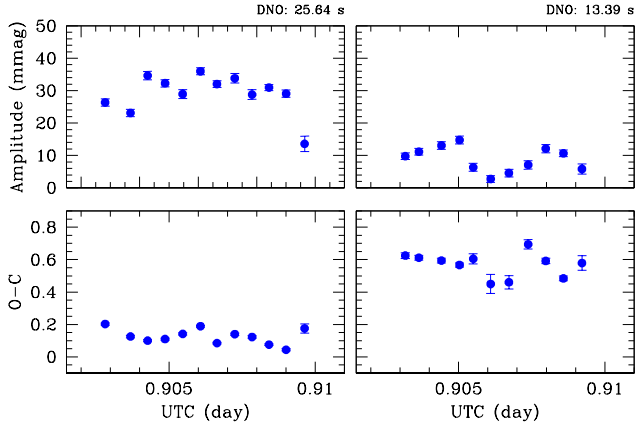,width=8.8cm}}}
  \caption{The $A - \phi$ diagrams of the first section of run SALT004 for the
25.64-s DNO (left panels: each dot represents $\sim$ 2 DNO cycles; there is a 33\% overlap.) 
and the 13.39-s DNO (right panels: each dot represents about $\sim$ 5 DNO cycles; there is a 50\%
overlap), respectively.}
 \label{SALT004omc2}
\end{figure}

  FTs of the sections of light curve shown in Fig.~\ref{SALT004lc1} show only fundamentals 
of the DNO modulations -- although the departure from sinusoidality of some of the 
pulsations shows that there must be harmonics present they are not detectable above the 
noise. These oscillations in VW Hyi are almost identical to those commonly observed in 
intermediate polars, for example see figure 1 of Warner \& Cropper (1984), where the light 
curve of V1223 Sgr has some sharp spikes in it similar to those in Fig.~\ref{SALT004lc1} but 
where it is also stated that no first harmonic is detectable in the FTs of the 
whole light curves.

\subsection{The 2008 January 6 Run (SALT017)}

\begin{figure}
\centerline{\hbox{\psfig{figure=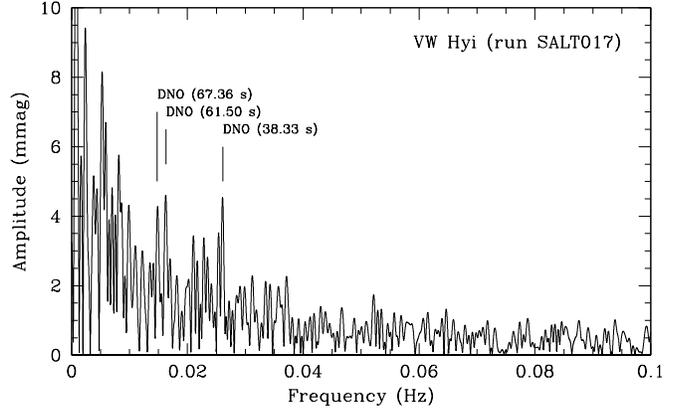,width=8.8cm}}}
  \caption{The Fourier transform of run SALT017 after prewhitening with the $\sim$ 770-s 
modulation.}
 \label{SALT017ft1}
\end{figure}

As mentioned above, there are no prominent DNOs in the lightcurve. The run is so 
short in length that we cannot certainly identify the nature of the three large 
amplitude peaks, with separations $\sim$ 770 s. After prewhitening (i.e. subtraction
from the light curve) with these modulations, 
the FT (Fig.~\ref{SALT017ft1}) shows several low amplitude oscillations that have significant 
relationships among their frequencies. The 5.7 mmag peak at 122.80 $\pm$ 0.27 s has 
a harmonic at 61.50 $\pm$ 0.08 s (4.4 mmag). The other two peaks are at 67.36 $\pm$ 0.10 s 
(4.1 mmag) and 38.33 $\pm$ 0.03 s (4.4 mmag). The latter is not a harmonic of the former, 
but if the former is a sidereal period the latter could be a harmonic of a corresponding 
synodic period with an implied (but not observed) QPO period $\sim$ 550 s 
(see discussion of similar structure in Sect.~3.2).

\subsection{The 2006 December 18 Run (SALT005)}

     FTs show that there are DNOs present about half of the time in this light curve, 
near three principal periods, viz 29 s, 42 s and 45 s. From the $A - \phi$ diagrams at 
these periods we selected for detailed analysis the regions where the amplitudes are 
largest (typically 5 mmag) or the phases are most coherent.  The individual coherent 
lengths range from 400 s to 740 s. The results are summarized in Table~\ref{dno7tab2}. 

\begin{figure}
\centerline{\hbox{\psfig{figure=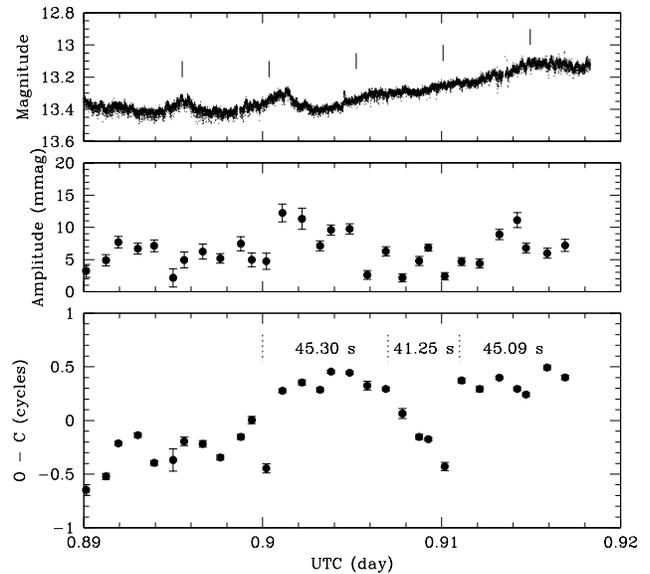,width=8.4cm}}}
  \caption{The A -- $\phi$ diagrams for the final section of run SALT005, relative
to a DNO at 45.30 s. Each dot represents $\sim$ 2 DNO cycles; there is a 50\% overlap.
The top panel shows the corresponding light curve with the 420-s QPO marked
by the vertical bars.} 
 \label{SALT005omc1}
\end{figure}

   We interpret the three periods in the following way: the 45 s and 29 s periodicities 
are an example of the 3:2 ratio first discussed in Paper IV (which here is a 3:2 ratio 
in the first harmonic of the synodic period), and this is supported by the brief 
appearance of a fundamental of the synodic period at 91 s in one section (see Table~\ref{dno7tab2}). 
The 45 s and 42 s periodicities are synodic and sidereal DNOs, for which we would 
expect a QPO at $\sim$ 460 s, and there is indeed a strong QPO at 420 s (see Fig.~\ref{SALT005omc1}). 
Also seen in Fig.~\ref{SALT005omc1} is the switch from synodic to sidereal period and back, as seen before 
in VW Hyi (see figure 12 of Paper IV -- which is switching of the second harmonic -- and 
seen in TY PsA in figures 1 and 2 in Paper VI). One section of the light curve contains 
the 29 s and 45 s modulations present simultaneously.

\section{Discussion}

In the various runs we have identified many of the DNO and QPO components seen in 
previous studies -- namely $\omega_{\rm DNO}$, $\omega_{\rm QPO}$, $\omega_{\rm DNO} - \omega_{\rm QPO}$, 
2 $\omega_{\rm DNO}$, 2 $\omega_{\rm QPO}$ and $\omega_{\rm lpDNO}$. In addition, for the first time 
we have seen the harmonic 2 ($\omega_{\rm DNO} - \omega_{\rm QPO}$) early in the evolution, when 
the fundamental period was $\sim$ 22 s, and the harmonics 2($\omega_{\rm DNO} - \omega_{\rm QPO}$) 
and 3($\omega_{\rm DNO} - \omega_{\rm QPO}$) at the very end of an outburst, when the fundamental 
period was $\sim$ 90 s. We will not attempt to explain these components beyond what has 
already been proposed in Papers II and IV, but we note the richness of the FTs, which 
is a challenge for any model.

\begin{figure}
\centerline{\hbox{\psfig{figure=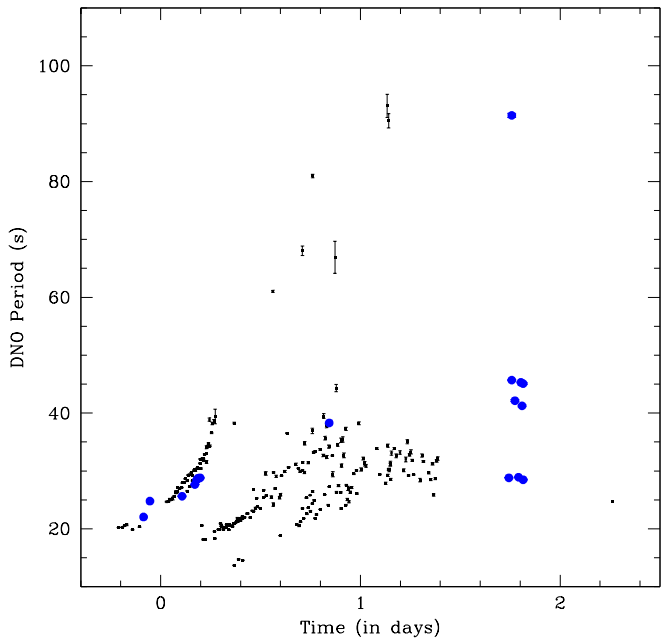,width=8.8cm}}}
  \caption{The evolution of DNO periods in VW Hyi after outburst. The big filled circles show
the DNOs reported in this paper.}
 \label{DNOevolSALT}
\end{figure}

   Our new observations add an interesting extension to the evolution of DNO periods and
 harmonics in VW Hyi. Fig.~\ref{DNOevolSALT} is based on figure 1 of Paper IV with addition of our new 
observations. We note first that our identification in Section 3.1 of the 24.78 s 
modulation as $P_{\rm SYN}$ rather than $P_{\rm DNO}$ results in better agreement with 
the $P_{\rm DNO}$ evolution. The DNOs observed at $T$ $\sim$ 1.8 d, and the later DNO mentioned in 
Sect. 2.2.4, demonstrate that after $T$ $\sim$ 1.1 d the DNOs cease to increase, as was 
suspected in Paper IV. The maximum value of the implied $P_{\rm DNO}$ is $\sim$ 90 s, which 
is essentially the same as the maximum value of $P_{\rm lpDNO}$ in VW Hyi (Paper III) -- the 
lpDNOs show little evolution in VW Hyi, increasing in period by about 10\% through 
an outburst, ultimately converging on the same period as that of the fundamental DNO.

\section*{Acknowledgments}

We thank Albert Jones for providing alerts of outbursts of VW Hyi, and Pauline Loader 
for supplying a file of the archived observations of VW Hyi collected by the Royal 
Astronomical Society of New Zealand. We kindly acknowledge the use of the AAVSO observational
archive of VW Hyi for determining the shape of the 2008 January outburst light curve.
PAW's and BW's research is supported by the 
University of Cape Town and the National Research Foundation of South Africa.
All of the observations reported in this paper were obtained with the Southern 
African Large Telescope (SALT), a consortium consisting of the National Research 
Foundation of South Africa, Nicholas Copernicus Astronomical Center of the Polish 
Academy of Sciences, Hobby Eberly Telescope Founding Institutions, Rutgers University, 
Georg-August-Universit\"{a}t G\"{o}ttingen, University of Wisconsin - Madison, Carnegie Mellon 
University, University of Canterbury, United Kingdom SALT Consortium, University of 
North Carolina - Chapel Hill, Dartmouth College, American Museum of Natural History 
and the Inter-University Centre for Astronomy and Astrophysics, India.


\begin{thebibliography}{99}
\bibitem{dod06}   O'Donoghue D., et al., 2006, MNRAS, 372, 151
\bibitem{p78}     Paczynski B., 1978, in Zytkow A., ed., Nonstationary Evolution of Close Binaries.
   Polish Scientific Publ., Warsaw, p.~89
\bibitem{pre06}   Pretorius M.L., Warner B., Woudt P.A., 2006, MNRAS, 368, 361 (paper V)
\bibitem{ste01}   Steeghs D., O'Brien K., Horne K., Gomer R., Oke J.B., 2001, MNRAS, 323, 484
\bibitem{vanA87}  van Amerongen S., Damen E., Groot M., Kraakman H., van Paradijs J., 1987, MNRAS, 225, 93
\bibitem{war95}   Warner B., 1995, Cataclysmic Variable Stars, Cambridge Univ. Press, Cambridge
\bibitem{war04}   Warner B., 2004, PASP, 116, 115
\bibitem{wab78}   Warner B., Brickhill A.J., 1978, MNRAS, 182, 777
\bibitem{wacr84}  Warner B., Cropper M., 1984, MNRAS, 206, 261
\bibitem{wapr08}  Warner B., Pretorius M.L., 2008, MNRAS, 383, 1469 (paper VI)
\bibitem{wawo02}  Warner B., Woudt P.A., 2002, MNRAS, 335, 84 (paper II)
\bibitem{wawo06}  Warner B., Woudt P.A., 2006, MNRAS, 367, 1562 (paper IV)
\bibitem{wwp03}   Warner B., Woudt P.A., Pretorius M.L., 2003, MNRAS, 334, 1193 (paper III)
\bibitem{wowa02}  Woudt P.A., Warner B., 2002, MNRAS, 333, 411 (paper I)
\end{thebibliography}
\end{document}